\documentclass[pra,showpacs]{revtex4}

\usepackage{amsmath,graphicx,epsfig,bm}
 \begin{document}
 
\title{Lowest Landau-level description of a Bose-Einstein condensate in a rapidly rotating anisotropic trap}
\author { Alexander L.\ Fetter}
\affiliation {Departments of Physics and Applied Physics, Stanford University,
   Stanford, CA 94305-4045, USA}

\date{\today}

\begin {abstract}
 A rapidly rotating  Bose-Einstein condensate in a symmetric two-dimensional trap can be  described with the lowest Landau-level  set of states.  In this case, the  condensate wave function $\psi(x,y)$ is a Gaussian function of $r^2 = x^2 + y^2$, multiplied by an analytic function $P(z) $ of the single complex  variable $z= x+ i y$; the zeros of $P(z)$ denote the positions of the vortices.  Here, a similar  description is used for  a rapidly rotating {\em anisotropic} two-dimensional trap with arbitrary anisotropy ($\omega_x/\omega_y\le 1$).  The corresponding condensate wave function $\psi(x,y)$ has the form  of a complex anisotropic Gaussian with a phase proportional to $ xy$, multiplied by an analytic function $ P(\zeta)$, where $\zeta \propto x + i \beta_- y$ and $ 0\le \beta_- \le 1$ is  a real   parameter that depends on the trap anisotropy and the rotation frequency.  The zeros of $P(\zeta)$ again fix the locations of the vortices.  Within the set of lowest Landau-level states  at zero temperature, an anisotropic parabolic density  profile provides an absolute minimum for the energy, with  the vortex density   decreasing  slowly and anisotropically away from the trap center.
\end{abstract}
\maketitle

 \section{Introduction}
 The experimental creation of rapidly rotating Bose-Einstein condensates (BEC) generally involves anisotropic rotating trap potentials~\cite{Ma00,Ab01,Ha01}, yet most  theoretical analyses of such systems have relied on an isotropic trap~\cite{Ho01,Ba04,Co04,Wa04,Af05,Wa06}.   As emphasized  by Ho~\cite{Ho01}, the low-lying states  in a symmetric  two-dimensional trap are closely analogous to those in the  lowest Landau level for a charged particle in a uniform magnetic field.   This analogy allows a simplified description in the limit that the rotation frequency $\Omega$ approaches the frequency $\omega_0$  of the symmetric confining trap.   If the typical interaction energy is small compared to the spacing $\approx 2\hbar\omega_0$ between adjacent Landau levels, then the condensate wave function $\psi$ of the interacting rotating BEC can be constructed as a linear combination of the lowest Landau-level (LLL) states. The $n$th  such state  is simply proportional to $z^n$, where $z =x+iy$, multiplied by the ground-state Gaussian.  It follows that such a linear combination involves an analytic function of $z$ that is usually approximated by a polynomial $P(z) \propto \prod_j(z-z_j)$, where the zeros $z_j$ of the polynomial represent the positions  of the vortices in the two-dimensional condensate.  If the vortex density is strictly uniform, then the overall density profile is also Gaussian with an effective condensate radius  that grows and ultimately diverges as $\Omega\to \omega_0$~\cite{Ho01}.  In fact, this system can lower its energy by slightly reducing the vortex density near the outer edge of the condensate, and the actual density profile has a quadratic shape (an inverted parabola)~\cite{Ba04,Co04,Wa04,Af05}, as shown by both analytical and numerical studies.  
 
 The quantum-mechanical problem of a particle in a rotating two--dimensional anisotropic harmonic trap is exactly soluble~\cite{Va56,Li01,Ok04}, although the corresponding eigenstates have not   been discussed previously in full detail.  We here amplify  Valatin's description~\cite{Va56} to construct the anisotropic analogs of the LLL states.  Each such state $\varphi_{n0}(x,y)$ involves  the anisotropic complex Gaussian ground-state eigenfunction $\varphi_{00}(x,y)$, multiplied by a polynomial $p_n(\zeta)$, where $\zeta\propto x+i\beta_- y$ is a single ``stretched'' complex variable, and $0 \le \beta_- \le 1 $ is  a real  parameter that depends on the trap anisotropy and the rotation frequency.  Thus a linear combination of these LLL states for an anisotropic trap again involves an analytic function of the single complex variable $\zeta$ (apart from the  common overall  factor $\varphi_{00}$). The corresponding zeros again represent the positions of the vortices, now in the rotating anisotropic BEC.  

Section II focuses on the eigenstates of the single-particle Hamiltonian $H_0$ for a rotating anisotropic harmonic potential, starting with the classical trajectories and then obtaining the explicit form of the low-lying quantum mechanical states $\varphi_{n0}$ that are the analogs of the lowest Landau-level states for a charged particle in a magnetic field.  As in that case, the expectation value of both $H_0$ and the angular momentum  $L_z$ in the lowest Landau level can be reduced to corresponding expectation values of $x^2$ and $y^2$.  The interacting dilute Bose-Einstein gas in this rapidly rotating anisotropic trap is treated in Sec.~III.  For a trial  state constructed as a linear combination of $\varphi_{n0}(x,y)$, an anisotropic parabolic density profile provides the absolute minimum of the energy.  The density of vortices is constant near the center but decreases slowly toward the edge of the condensate.  Section IV contains a discussion and suggestions for additional study.

\section{single-particle eigenstates}

Consider a particle of mass $m$ in an anisotropic two-dimensional harmonic potential (for definiteness, I assume  oscillator frequencies $\omega_x \le \omega_y$) that rotates uniformly at an angular velocity $\bm \Omega = \Omega\,\hat{\bm z}$ perpendicular to the plane of the motion.  In the rotating frame, this potential is time-independent,   with the  Hamiltonian  
\begin{equation} \label{h0}
H_0=\frac{p_x^2+p_y^2}{2m} + \frac{1}{2} m\left(\omega_x^2\,x^2 + \omega_y^2 \,y^2\right) -\Omega\left(xp_y-yp_x\right),
\end{equation}
where  the last factor involves the angular momentum $L_z = xp_y-yp_x$.  If  $\omega_x < \omega_y$, the centrifugal force preferentially expands the condensate along the $x$ axis.   Although this case ($\omega_x < \omega_y $) is of principal interest here,  it will also be  valuable to see how the more familiar symmetric  case emerges in the limit  $\omega_x =\omega_y =\omega_0$.  

\subsection{Classical dynamical trajectories}

The normal modes of Eq.~(\ref{h0}) are readily determined to have the frequencies~\cite{La23,Va56,Li01,Ok04,Pi03} 
\begin{equation} \label{opm}
\omega_\pm^2 = \omega_\perp^2 + \Omega^2 \mp\sqrt{\textstyle{\frac{1}{4}} \left(\omega_y^2-\omega_x^2\right)^2 + 4\omega_\perp^2\,\Omega^2},
\end{equation}
where $\omega_\perp^2 = \frac{1}{2}\left(\omega_x^2 + \omega_y^2\right)$ is the mean-squared oscillator frequency.   This general result contains several important limits.
\begin{enumerate}
\item In the symmetric case $\omega_x=\omega_y =\omega_0$, the plus (minus) modes have frequencies $\omega_\pm = \omega_0\mp \Omega$.  Specifically, the plus mode with frequency $\omega_+=\omega_0-\Omega $ has a {\it reduced\/} frequency when viewed from the rotating frame (as is evident physically) and a positive angular momentum (which explains the notation).  Correspondingly, the minus mode has an {\it increased\/} frequency $\omega_- = \omega_0 + \Omega$ and a negative angular momentum.
\item If $\omega_x < \omega_y$, then the modes are nondegenerate even for $\Omega = 0$, when they reduce to $\omega_+ = \omega_x$ and $\omega_- = \omega_y$.
\item For an anisotropic trap ($\omega_x < \omega_y$) and rapid rotation with  $\delta = 1 -  \Omega/\omega_x\to 0^+$,  the plus normal-mode frequency vanishes, with 
\begin{equation} \label{op}
\omega_+^2 \approx \frac{2\omega_x^2 \left(\omega_y^2- \omega_x^2\right)}{3\omega_x^2 + \omega_y^2}\,\delta.
\end{equation}
Thus $-\partial \omega_+/\partial \Omega$ diverges for small $\delta$ like $\delta^{-1/2}$.
In contrast,  the minus  normal-mode frequency remains finite at $\delta = 0$, with  
\begin{equation}\label{om}
\omega_-^2\approx 3\omega_x^2 + \omega_y^2. 
\end{equation}
\item In the case of  a rapidly rotating nearly symmetric trap with two small parameters~\cite{Ok04} $\delta = 1-\Omega/\omega_x$ and $\eta = \omega_y/\omega_x -1$, these eigenfrequencies simplify to 
 \begin{equation}\label{freq}
\frac{\omega_+}{\omega_x} \approx \sqrt{\delta(\eta + \delta)},\qquad\frac{ \omega_-}{\omega_x}  \approx 2 + {\textstyle \frac{1}{2}}\eta -\delta.
\end{equation}
Note the sensitivity  of $\omega_+$ in Eq.~(\ref{freq})  to the order of  limiting procedures: (i) first $\eta = 0$ (namely $\omega_x = \omega_y = \omega_0 $) and then $\delta \to 0$ or (ii)  $\delta \to 0$ (namely $\Omega \to \omega_x$) at fixed $\omega_x < \omega_y$. For a symmetric trap, the plus frequency $\omega_+$  vanishes linearly with the small parameter $\delta$; in contrast, for the anisotropic trap with $\omega_x < \omega_y$, the plus frequency vanishes like $\sqrt \delta$, with a  coefficient proportional to $\sqrt{\eta}$.  
\end{enumerate}
Figure 1(a) shows the two different normal-mode frequencies $\omega_\pm$ (normalized to $\omega_x$)  as functions of $\Omega/\omega_x$ for the typical anisotropy  $\omega_y/\omega_x = 1.2$.

  \begin{figure}[ht] 
  \includegraphics[width=4in]{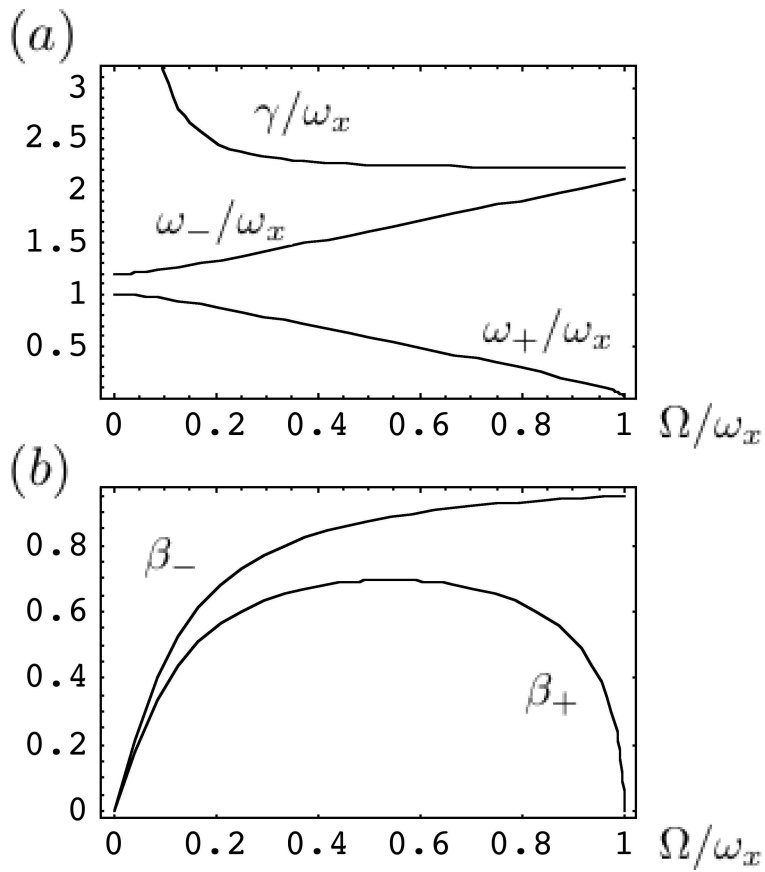}
\vspace{.2in}

 \caption{Behavior of relevant dimensionless quantities as a function of dimensionless rotation speed $\Omega/\omega_x$ for the typical anisotropy  $\omega_y/\omega_x = 1.2$.  (a) dimensionless normal mode frequencies $\omega_+/\omega_x$ and $\omega_-/\omega_x$ and dimensionless auxiliary frequency $\gamma/\omega_x$;  (b) dimensionless polarization parameters $\beta_+$ and $\beta_-$.}
 \label{fig1}
 \end{figure}

For a  symmetric  trap with $\omega_x = \omega_y =\omega_0$, the plus (minus) normal modes with frequencies $\omega_\pm  =\omega_0 \mp \Omega$  have counterclockwise (clockwise) circular orbits with positive (negative) helicity.  To understand the physics  of the two normal modes in the more general anisotropic case   $\omega_x < \omega_y$, consider first the  motion in the  plus mode~\cite{Va56}.  It has the form 
\begin{eqnarray}\label{xp}
x_+(t)  =  x_0 e^{-i\omega_+ t},\qquad 
y_+(t)  =  i\beta_+ x_0  e^{-i\omega_+ t}, 
\end{eqnarray}
where $0\le \beta_+\le 1 $ is a  real non-negative dimensionless parameter and   $x_0$ is an arbitrary amplitude.  A detailed analysis shows that the  parameter $\beta_+$ has two alternative representations
\begin{equation} \label{bp}
\beta_+ = \frac{\omega_x^2 -\omega_+^2-\Omega^2}{2\Omega\omega_+} =  \frac{2\Omega\omega_+} {\omega_y^2 -\omega_+^2-\Omega^2}\,.
\end{equation}
In the rotating frame, the plus orbit is an ellipse with major axis oriented along $\hat{\bm x}$;  the polarization  parameter $\beta_+$  gives the ratio of the minor to major axes. The second form of Eq.~(\ref{bp}) shows that $\omega_+/\beta_+\approx (\omega_y^2-\omega_x^2)/(2\omega_x)$ for $\delta \to  0$, so that $\omega_+$ and $\beta_+$ both vanish in this limit like $\sqrt{\delta}$.  Physically, this behavior reflects the rotation-induced cancellation of the harmonic confinement in the $\hat{\bm x}$ direction;  the orbit then becomes linearly polarized as $\Omega\to\omega_x$.   The plus  motion is counterclockwise, with  positive helicity  and   positive angular momentum.

Similarly, the orbit for the minus mode has the parametric representation  
 \begin{eqnarray}\label{xm}
x_-(t)  =  i\beta_-y_0 e^{-i\omega_- t},\qquad 
y_-(t)  = y_0  e^{-i\omega_- t}, 
\end{eqnarray}
where  $y_0$ is an arbitrary amplitude and $0\le \beta_-\le 1 $ is a  real non-negative parameter with   two alternative representations~\cite{Va56}
\begin{equation} \label{bm}
\beta_- = \frac{\omega_-^2 -\omega_y^2+\Omega^2}{2\Omega\omega_-} =  \frac{2\Omega\omega_-} {\omega_-^2 -\omega_x^2+\Omega^2}\,.
\end{equation}
Unlike $\beta_+$, these relations show that $\beta_-$ has a nonzero  limit $\beta_-\approx 2\omega_x/(3\omega_x^2 + \omega_y^2)^{1/2}\le 1$ as $\delta \to 0$.  
In the rotating frame, the minus orbit is an ellipse with major axis oriented along $\hat{\bm y}$;  the polarization parameter  $\beta_-$  gives the ratio of the minor to major axes.  The minus  motion is clockwise, with  negative helicity and  negative angular momentum. 
  
It is again instructive to specialize these results to the case of a rapidly rotating nearly symmetric trap~\cite{Ok04},  making use of  the small parameters $\delta = 1 - \Omega/\omega_x$ and $\eta = \omega_y/\omega_x - 1$.  A straightforward expansion yields the approximate  polarization parameters
\begin{equation}\label{beta}
\beta_+ \approx \sqrt{\frac{\delta}{\eta + \delta}},\qquad \beta_- \approx 1 - {\textstyle \frac{1}{4}}\eta.
\end{equation}
Although  $\beta_-$ varies  smoothly in this limit,  the corresponding  polarization $\beta_+$ for the positive mode has a   singular behavior that depends on the relative magnitude of  the two small parameters.  More generally,  $\beta_+$ and $\beta_-$ both  have the value 1 in the limit of a symmetric  trap, independent of $\Omega$, since the resulting motion is circularly polarized.  Figure~1(b)  shows the dependence of the two parameters $\beta_\pm$ on $\Omega/\omega_x$ for an anisotropy $\omega_y/\omega_x = 1.2$.  Note the singular slope of $\omega_+$ and $\beta_+$ near the upper limit.

\subsection{Bogoliubov canonical transformation to diagonal Hamiltonian}

 The structure of $H_0$ in Eq.~(\ref{h0}) is unusual because  the term $\Omega L_z = \Omega\left(yp_x-xp_y\right)$ couples the otherwise independent $x$ and $y$ motions. This situation can be clarified by introducing the conventional ladder operators~\cite{Sh94}
 \begin{equation}\label{adag}
a_x = \frac{1}{\sqrt 2}\left(\frac{x}{d_x} + i\frac{d_x p_x}{\hbar}\right),\quad a_x^\dagger= \frac{1}{\sqrt 2}\left(\frac{x}{d_x} - i\frac{d_x p_x}{\hbar}\right),
\end{equation}
where $d_x = \sqrt{\hbar/(m\omega_x)}$, and similarly for $a_y$ and $a_y^\dagger$.  With these operators, it is straightforward to see that  the term $\Omega L_z$ is  proportional to~\cite{Li01} 
\begin{equation}
i\Omega\left[\left(\omega_y + \omega_x\right)\left(a_x^\dagger a_y - a_y^\dagger a_x\right) + \left(\omega_y - \omega_x\right)\left(a_x a_y - a_y^\dagger a_x^\dagger\right) \right].
\end{equation}
The first term is ``diagonal'' in the creation and annihilation operators, but the second is ``off-diagonal,'' similar to Bogoliubov's approximate Hamiltonian for  a dilute Bose-Einstein gas~\cite{Bo47}.  Unfortunately, a direct diagonalization  based on these ``particle''  operators involves considerable algebraic complexity~\cite{Li01,Ok04}.  

Thus, it is preferable to return to the original single-particle Hamiltonian in Eq.~(\ref{h0}).  Since $H_0$   is quadratic in the coordinates and momenta, it can be diagonalized  with a canonical transformation to new variables that obey the same Poisson brackets (in the classical case) or the same commutators (in the quantum case). 
Specifically,  I  follow Valatin~\cite{Va56} and introduce the generating function 
\begin{equation}\label{S}
S(x,y;Q_+,Q_-) = -m\gamma\left[\lambda_+\lambda_-Q_+Q_-+ \textstyle{\frac{1}{2}}\left(\lambda_+^2 + \lambda_-^2\right) xy -\lambda_+Q_+y-\lambda_-Q_-x\right].
\end{equation}
Here, $ Q_\pm$ are new canonical coordinates, $\lambda_\pm$ are dimensionless constants given by
 \begin{equation}\label{lam}
\lambda_\pm^2 =\frac{\omega_\pm}{\mu_\pm}\, , \quad\hbox{with} \quad \mu_\pm = \omega_\pm + \beta_+\beta_-\omega_\mp,
\end{equation}
and $\gamma=\mu_+/\beta_+ = \mu_-/\beta_-$ has the dimensions of a frequency.
It follows  from Eqs.~(\ref{opm}), (\ref{bp}), (\ref{bm}) and  (\ref{lam}) that  $\gamma$ has various equivalent representations
%\begin{equation}\label{gam}
%\gamma =\frac{\omega_+}{\beta_+} + \omega_-\beta_- = \frac{\omega_-}{\beta_-} + \omega_+\beta_+
%= \frac{\omega_-^2-\omega_+^2}{2\Omega} = \frac{\sqrt{\textstyle{\frac{1}{4}} \left(\omega_y^2-\omega_x^2\right)^2 + 4\omega_\perp^2\,\Omega^2}}{\Omega}.
%\end{equation}
\begin{eqnarray}\label{gam}
\gamma &=&\frac{\omega_+}{\beta_+} + \omega_-\beta_- = \frac{\omega_-}{\beta_-} + \omega_+\beta_+
\nonumber \\[.2cm]
&=& \frac{\omega_-^2-\omega_+^2}{2\Omega} = \frac{\sqrt{\textstyle{\frac{1}{4}} \left(\omega_y^2-\omega_x^2\right)^2 + 4\omega_\perp^2\,\Omega^2}}{\Omega}.
\end{eqnarray}
For a symmetric trap with $\omega_x=\omega_y =\omega_0$, this frequency reduces to $\gamma = 2\omega_0$ for all $\Omega$.  
In contrast, for an anisotropic trap,  $\gamma$ diverges as $\Omega$ becomes small and approaches the value $\approx (3\omega_x^2 +\omega_y^2)/(2\omega_x) > 2\omega_x$ for $\Omega\to \omega_x$.  
Figure 1(a) shows the normalized parameter $\gamma/\omega_x$ as a function of $\Omega/\omega_x$ for $\omega_y/\omega_x = 1.2$.

According to the general theory of  classical Hamiltonian dynamics~\cite{Fe03}, {\em any}  function like $S(x,y;Q_+,Q_-)$ that depends on both the old and new coordinates  will automatically generate a canonical transformation from old canonical variables to new  canonical variables, with the corresponding momentum variables given by 
\begin{eqnarray}\label{mom}
p_x = \frac{\partial S}{\partial x},&\quad&p_y=\frac{\partial S}{\partial y}, \nonumber \\[.2cm]
P_+ = -\frac{\partial S}{\partial Q_+},&\quad&P_-=-\frac{\partial S}{\partial Q_-}.
\end{eqnarray}
The first set of equations immediately yields the relations
\begin{eqnarray}\label{Q}
Q_+ & = & \left(\frac{\lambda_+^2+\lambda_-^2}{2\lambda_+}\right)x +\frac{p_y}{m\gamma\lambda_+},\nonumber\\[.2cm]
Q_- & = & \left( \frac{\lambda_+^2+\lambda_-^2}{2\lambda_-}\right)y +\frac{p_x}{m\gamma\lambda_-}
\end{eqnarray}
 that express the new coordinates $Q_\pm$ as linear combinations of the original coordinates and momenta.  Similarly, the second set of equations can be used to find the corresponding relations for the new momenta $P_\pm$
 \begin{eqnarray}\label{P}
P_+ & = & m\gamma\lambda_+\left(\frac{\lambda_+^2+\lambda_-^2}{2} - 1\right) y +\lambda_+p_x, \nonumber\\[.2cm]
P_- & = & m\gamma\lambda_-\left(\frac{\lambda_+^2+\lambda_-^2}{2} - 1\right) x +\lambda_-p_y.
\end{eqnarray}
For future reference, note  the following alternative relations 
\begin{eqnarray}\label{xy}
x & = & \lambda_+Q_+ -\frac{P_-}{m\gamma\lambda_-}, \nonumber \\[.2cm]
y & = & \lambda_-Q_- -\frac{P_+}{m\gamma\lambda_+};
\end{eqnarray}
they express the original coordinates in terms of the new canonical variables and will be valuable  in the subsequent analysis.

It is now straightforward to verify that the  Hamiltonian has the following simple diagonal form when expressed in the new canonical variables
\begin{equation}\label{hdiag}
H_0 = \frac{P_+^2}{2m} +\frac{1}{2} m\omega_+^2 Q_+^2  + \frac{P_-^2}{2m} +\frac{1}{2} m\omega_-^2 Q_-^2.
\end{equation}
One  strategy  is to substitute Eqs.~(\ref{Q}) and (\ref{P}) directly into (\ref{hdiag}), which eventually reproduces the original Eq.~(\ref{h0}).  This new Hamiltonian (\ref{hdiag}) has the great advantage of  immediately providing a quantum description of two independent harmonic oscillators with mass $m$ and frequencies $\omega_\pm$.  The corresponding    quantum-mechanical annihilation  operators $\alpha_\pm$ and creation operators $\alpha_\pm^\dagger$ follow from general quantum theory~\cite{Sh94} 
\begin{equation}\label{alpha}
\alpha_\pm = \frac{1}{\sqrt 2}\left(\frac{Q_\pm}{d_\pm} + i\frac{d_\pm P_\pm}{\hbar}\right),\quad \alpha_\pm^\dagger = \frac{1}{\sqrt 2}\left(\frac{Q_\pm}{d_\pm} - i\frac{d_\pm P_\pm}{\hbar}\right),
\end{equation}
where $d_\pm = \sqrt{\hbar/(m\omega_\pm)}$ are the oscillator lengths for the two separate modes.  These operators  obey the usual commutation relations $[\alpha_\pm,\alpha_\pm^\dagger] = 1$ (all other commutators vanish).  Note that $d_+$ diverges as $\Omega\to \omega_x$, whereas $d_-$ remains finite in the same limit.   In terms of these operators, the Hamiltonian takes the  form~\cite{Li01} 
\begin{equation}\label{hop}
H_0 = \textstyle{\frac{1}{2}}\hbar\omega_+ \left( \alpha_+^\dagger\alpha_+ + \alpha_+\alpha_+^\dagger\right) + \textstyle{\frac{1}{2}}\hbar\omega_- \left( \alpha_-^\dagger\alpha_- + \alpha_-\alpha_-^\dagger\right). \end{equation}

\subsection{Lowest Landau-level single-particle states for rotating anisotropic trap}

The single-particle ground state $\varphi_{00}$ is given by the  prescription $\alpha_\pm \varphi_{00} =  0$, which leads to the explicit representation 
\begin{equation}\label{gndst}
\varphi_{00} \propto \exp\left(-\frac{Q_+^2}{2d_+^2} -\frac{Q_-^2}{2d_-^2} \right)
\end{equation}
as a Gaussian function of the two new coordinates $Q_\pm$. Correspondingly, the complete set of normalized   single-particle states $\varphi_{n_+n_-}$ is specified by two non-negative integers $n_\pm$
\begin{equation}\label{ex}
\varphi_{n_+n_-} = \frac{(\alpha_+^\dagger)^{n_+}}{\sqrt{n_+!}}\,\frac{(\alpha_-^\dagger)^{n_-}}{\sqrt{n_-!}}\,\varphi_{00}.
\end{equation}
Here, the eigenstate state $\varphi_{n_+n_-}$ has $n_+ \>(n_-)$ quanta with frequency $\omega_+\> (\omega_-)$;  the   energy eigenvalue  is
\begin{equation} \label{epsilon}
\epsilon_{n_+n_-} = \hbar\omega_+\left(n_+ + \textstyle{\frac{1}{2}} \right) + \hbar\omega_-\left(n_- + \textstyle{\frac{1}{2}}\right).
\end{equation}
 This eigenstate $\varphi_{n_+n_-} $ has an  angular momentum~\cite{Li01} 
 \begin{equation}\label{L}
 L_{n_+n_-} = -\frac{\partial \epsilon_{n_+n_-}}{\partial \Omega}  = -\hbar\frac{\partial \omega_+}{\partial \Omega}\left(n_+ + \frac{1}{2} \right) - \hbar\frac{\partial \omega_-}{\partial \Omega}\left(n_- + \frac{1}{2} \right).
 \end{equation}
 Since $\partial\omega_+/\partial \Omega $ is negative ($\partial \omega_-/\partial \Omega$ is positive), this result confirms that the plus (minus) mode has positive (negative) angular momentum.  For the special case of a symmetric trap, the detailed form of these eigenstates is well known~\cite{Co77,Li99}. 
 
 It is important to re-express the ground-state wave function $\varphi_{00}$ in terms of the original canonical coordinates $x$ and $y$.  Valatin~\cite{Va56} uses the generating function $S(x,y;Q_+,Q_-)$ in Eq.~(\ref{S}) to obtain the explicit (factorized) expression 
 \begin{eqnarray}\label{phi00}
\varphi_{00}(x,y)  \propto  \exp\left[-\frac{m\gamma \left(\beta_+ x^2 +\beta_- y^2\right)}{2\hbar\left(1+\beta_+\beta_-\right)}\right] \,\exp\left\{ i \frac{m\,xy}{\hbar}\left[\frac{\gamma}{1+\beta_+\beta_-}-\frac{1}{2}\left(\frac{\omega_+}{\beta_+}+\frac{\omega_-}{\beta_-}\right)\right]\right\}.
\end{eqnarray}
Each of these two factors has an interesting structure.
\begin{enumerate}
\item The first factor   is a real  anisotropic Gaussian with characteristic lengths $a_x$ and $a_y$ given by 
\begin{equation}\label{a}
a_x^2 =\frac{1+\beta_+\beta_-}{\beta_+}\,\frac{\hbar}{m\gamma},\qquad a_y^2 =\frac{1+\beta_+\beta_-}{\beta_-}\,\frac{\hbar}{m\gamma}.
\end{equation}
  In the limit of a  symmetric trap,  this Gaussian ground state becomes $\varphi_{00}(x,y) \propto \exp\left[-m\omega_0\left(x^2+y^2 \right)/(2\hbar)\right] $,  with the  expected oscillator length $d_0=\sqrt{\hbar/(m\omega_0)}$.
More generally, the ground-state density  for a rotating anisotropic trap is an anisotropic Gaussian, given by  the corresponding normalized wave function 
\begin{equation}\label{dens00}
\left|\varphi_{00}(x,y\right)|^2 = \frac{1}{\pi a_xa_y}\,\exp\left(-\frac{x^2}{a_x^2} -\frac{y^2}{a_y^2}\right).
\end{equation}
For a rapidly rotating  anisotropic trap ($\omega_x <\omega_y$ and $\delta = 1-\Omega/\omega_x\to 0$),   the length  $a_x$ diverges because $\beta_+\to 0$, but $a_y$ remains finite. In this limit, the ground-state density becomes an essentially one-dimensional strip with Gaussian  transverse profile and finite width $a_y$~\cite{Si05,Sa05}.
\item The second factor of $\varphi_{00}$ involves  a complex phase   proportional to $xy$, which reflects the irrotational flow induced by the rotating anisotropic trap~\cite{La45,Fe74,Pi03a}. 
  The  factor in square brackets (the coefficient of $imxy/\hbar$) has a rather intricate structure.  It vanishes for a symmetric trap, because   $\beta_\pm = 1$ and $\omega_+ +\omega_- =\gamma =  2\omega_0$.  It also vanishes for a stationary anisotropic trap, but this limit requires a detailed analysis because each term separately diverges as $\Omega \to 0$.  This phase will be seen to play an essential role in the following construction of the lowest Landau-level states.
\end{enumerate}

With Eqs.~(\ref{Q}) and (\ref{P}), the  operators $\alpha_\pm$ and $\alpha_\pm^\dagger$ defined in Eqs.~(\ref{alpha}) are readily expressed in terms of the original coordinates $x,y$ and momenta $p_x,p_y$.  It is not hard to  verify explicitly that $\alpha_\pm\,\varphi_{00}(x,y) = 0$.
The more interesting question is the form of the low-lying states $\varphi_{n0}(x,y)$, which are the  analogs of the lowest Landau-level states but now  for an anisotropic  rotating trap.  In this case, the state $\varphi_{n0}$ has $n$ quanta of the plus mode (whose frequency  $\omega_+$ becomes small  $\propto \sqrt\delta$ for $\delta = 1-\Omega/\omega_x \to 0$), and zero  quanta of the minus mode (which has a finite frequency $\omega_-$ in the same limit).  

The basic relation $\alpha_+\varphi_{00} = 0$ implies that $(Q_+/\sqrt2 \,d_+)\varphi_{00} = -i (d_+P_+/\sqrt2 \,\hbar)\varphi_{00}$.  Thus $\alpha_+^\dagger \varphi_{00} = (\sqrt 2\, Q_+/d_+)\varphi_{00} =  -i (\sqrt2 \,d_+P_+/ \hbar)\varphi_{00}$, and   a  straightforward calculation yields
\begin{equation}\label{phi10}
\varphi_{10}(x,y) =\alpha_+^\dagger \,\varphi_{00}(x,y) = \zeta\,\varphi_{00}(x,y),
\end{equation}
where $\varphi_{00}$ is the normalized ground state, and 
\begin{eqnarray}\label{zeta}
\zeta &=& \frac{\sqrt 2\left(x+i\beta_-y\right)}{d_+\lambda_+\left(1+\beta_+\beta_-\right)} = \sqrt{\frac{2m\gamma\beta_+}{ \hbar}}\,\frac{x+i\beta_-y}{1+\beta_+\beta_-} \nonumber \\[.2cm]
&= &\sqrt\frac{2}{1+\beta_+\beta_-}\,\frac{x+i\beta_-y}{a_x}
\end{eqnarray}
is a dimensionless complex variable involving a ``stretched'' combination $x+ i \beta_- y$.  As shown in Fig.~1(b), the polarization parameter $\beta_-$ is real and less that 1 in the limit of rapidly rotating anisotropic trap.   This complex variable   reduces to~\cite{Ho01} $\zeta = (x+iy)/d_0$ for a rotating symmetric  trap, where $d_0 = \sqrt{\hbar/(m\omega_0)}$.  In the more general case of a rotating anisotropic trap, the characteristic length that appears in (\ref{zeta}) is essentially $a_x$ from Eq.~(\ref{a}), apart from the common factor $\sqrt{\frac{1}{2}(1+\beta_+\beta_-)}$ and the additional factor $\beta_-$ for $y$;  both these factors remain finite as $\Omega\to \omega_x$.  This quasi-isotropic behavior for $\zeta$ is very different from the anisotropy seen in the two lengths $a_x$ and $a_y$ that determine the $x$ and $y$ structure of the ground-state density  $|\varphi_{00}|^2$.

The higher states within the lowest Landau level have a similar structure.  For example, $\varphi_{20} = (\alpha_+^\dagger)^2\,\varphi_{00}/\sqrt 2$ can be written as  
\begin{eqnarray}
\varphi_{20}(x,y) &=& \frac{\alpha_+^\dagger}{\sqrt 2} \,\varphi_{10}(x,y) = \frac{\alpha_+^\dagger}{\sqrt 2} \,\zeta\,\varphi_{00}(x,y) \nonumber \\[.2cm]
&=& \frac{1}{\sqrt 2}\left(-\left[\zeta,\alpha_+^\dagger\right] + \zeta^2\right)  \varphi_{00}(x,y).
\end{eqnarray}
The commutator is readily evaluated with Eqs.~(\ref{alpha}), (\ref{Q}), (\ref{P}) and (\ref{zeta}), yielding 
\begin{equation}\label{c}
\left[\zeta,\alpha_+^\dagger\right]  =\frac{1-\beta_+\beta_-}{1+\beta_+\beta_-}  \equiv c,
\end{equation}
which defines the constant $c$ (it depends on the trap frequencies $\omega_x$, $\omega_y$ and the rotation speed $\Omega$).  In general, $0\le c\le 1$, but  it vanishes identically for a symmetric trap since $\beta_\pm =1 $ in this case.  Thus $\varphi_{20} = (\zeta^2-c)\,\varphi_{00}/\sqrt 2$, namely an even polynomial in $\zeta$ times the complex Gaussian ground state  $\varphi_{00}$.  As a check on this analysis, note that $[\alpha_+,\alpha_+^\dagger]\varphi_{00}= \varphi_{00}$, and direct calculation verifies that $[\alpha_+,\zeta] = 1$.

The general lowest Landau-level state follows from similar arguments [it is essential here that the commutator 
  (\ref{c}) is a pure number, independent of $x$ and $y$]
  \begin{equation} \label{phin}
\varphi_{n0}(x,y) = \frac{1}{\sqrt{n!}}\,p_n(\zeta)\,\varphi_{00}(x,y),
\end{equation}
where $p_n(\zeta)$ is a polynomial of order $n$ that obeys the symmetry condition $p_n(-\zeta) = (-1)^n\,p_n(\zeta)$.  These polynomials  are easily obtained recursively with the relation
\begin{equation}\label{pn}
p_{n+1}(\zeta) = \zeta\,p_n(\zeta) - c\,\frac{dp_n(\zeta)}{d\zeta},
\end{equation}
with the  first few given explicitly as $p_0 = 1$, $p_1 = \zeta$, $p_2 = \zeta^2 -c$, $p_3 = \zeta^3 - 3c\zeta, \>\cdots$.  The Hermite polynomials $H_n(x)$ obey a similar recursion relation~\cite{Wo} $H_{n+1}(x) = 2x\,H_n(x) -H_n'(x)$.  Direct comparison shows that  
\begin{equation}\label{poly}
p_n(\zeta) = \left(\frac{c}{2}\right)^{n/2} H_n\left(\frac{\zeta}{\sqrt{2c}}\right),
\end{equation}
which readily reproduces the explicit forms given above for small $n = 0, \cdots, 3$.  Oktel~\cite{Ok04}  obtained an analogous but less general result in the  special limit of small anisotropy and rapid rotation.  For a symmetric trap (with $\beta_\pm = 1$ and $c = 0$),   it   follows directly that $p_n(\zeta)$ reduces to  the $n$th power   of $(x+ i y)/d_0$.   

\subsection{Expectation value of  single-particle  $H_0$  for general lowest Landau-level state}

Let $\psi_{LLL} = \sum_n c_n\varphi_{n0}$ be a general linear combination of lowest Landau-level states $\{\varphi_{n0}\}$, normalized  with the condition $\int dxdy \,|\psi_{LLL}|^2 = 1$.  The expectation value of $H_0$ in Eq.~(\ref{h0}) is given by the equivalent  operators in (\ref{hop})
\begin{eqnarray}
\langle H_0 \rangle &=& \int dxdy\,\psi_{LLL}^*\left[ \textstyle{\frac{1}{2}}\hbar\omega_+ \left( \alpha_+^\dagger\alpha_+ + \alpha_+\alpha_+^\dagger\right) + \textstyle{\frac{1}{2}}\hbar\omega_- \left( \alpha_-^\dagger\alpha_- + \alpha_-\alpha_-^\dagger\right) \right]\psi_{LLL},
\end{eqnarray}
where the angular brackets denote the expectation value with the state $\psi_{LLL}$.
Since $\alpha_-\psi_{LLL}$ vanishes (by construction), this quantity reduces to 
\begin{equation}\label{h0LLL}
\langle H_0 \rangle = \textstyle{\frac{1}{2}}\hbar\omega_-  + \textstyle{\frac{1}{2}}\hbar\omega_+ \int dxdy\,\psi_{LLL}^* \left( \alpha_+^\dagger\alpha_+ + \alpha_+\alpha_+^\dagger\right)\psi_{LLL},
\end{equation}
where the first term is just the zero-point energy of the unoccupied minus mode.  For a symmetric trap, this expectation value is readily expressed in terms of the expectation value  $\langle x^2 + y^2\rangle $~\cite{Ho01,Af05}.  As shown below,  a similar but more intricate result holds for the rotating anisotropic trap.  

It is convenient to start from Eqs.~(\ref{xy}) that express $x$ and $y$ in terms of the new canonical variables $Q_\pm$ and $P_\pm$. In turn, these operators  are simply linear combinations of the corresponding oscillator variables $\alpha_\pm^\dagger$ and $\alpha_\pm$, as follows from (\ref{alpha}). For example, 
\begin{equation}
x = \frac{d_+\lambda_+}{\sqrt 2} \left(\alpha_+ + \alpha_+^\dagger\right) -\frac{\hbar}{\sqrt 2 i d_-\lambda_- m \gamma} \left(\alpha_ - -  \alpha_-^\dagger\right).
\end{equation}
The expectation value of $x^2$ is then given by 
\begin{equation}\label{x2}
\langle x^2 \rangle = \frac{\hbar }{2m\gamma \beta_+} \langle \alpha_+^\dagger\alpha_+ + \alpha_+\alpha_+^\dagger \rangle + \frac{\hbar }{2m\gamma \beta_+} \langle( \alpha_+^\dagger)^2  + \left(\alpha_+\right)^2 \rangle +\frac{\hbar \beta_-}{2m\gamma},
\end{equation}
where the cross-terms between plus and minus operators vanish because $\langle \alpha_-\rangle = \langle \alpha_-^\dagger \rangle= 0 $, and I use the relations $d_\pm^2\lambda_\pm^2 = (\hbar/m\omega_\pm)\,(\omega_\pm/\mu_\pm) = \hbar/m\gamma \beta_\pm$.
A similar calculation gives 
\begin{equation}\label{y2}
\langle y^2 \rangle = \frac{\hbar\beta_+ }{2m\gamma} \langle \alpha_+^\dagger\alpha_+ + \alpha_+\alpha_+^\dagger \rangle - \frac{\hbar\beta_+ }{2m\gamma } \langle( \alpha_+^\dagger)^2  + \left(\alpha_+\right)^2 \rangle +\frac{\hbar }{2m\gamma\beta_-},
\end{equation}
and an appropriate linear combination leads to the quantity in Eq.~(\ref{h0LLL}).  In this way, the desired  LLL expectation value $\langle H_0 \rangle$ has the simple form 
\begin{eqnarray}\label{h0LLLa}
\langle H_0\rangle & = & \frac{1}{2} \hbar\omega_- - \frac{1}{4}\hbar \omega_+ \left(\beta_+\beta_- + \frac{1}{\beta_+\beta_-}\right)
+\frac{1}{2} m\gamma \omega_+\left(\beta_+ \langle x^2 \rangle + \frac{1}{\beta_+} \langle y^2 \rangle \right).
\end{eqnarray}
For a symmetric trap with $\omega_x = \omega_y = \omega_0$, this result reduces to the familiar LLL expression $\langle H_0\rangle  = \hbar\Omega + m\omega_0\left(\omega_0 - \Omega \right) \langle r^2 \rangle$~\cite{Ho01,Af05}, where $r^2 = x^2 + y^2 $.

In the  case of a symmetric condensate, the special properties of the  LLL states yield   a  simple   well-known relation between the expectation value of the angular momentum  and the mean-squared radius~\cite{Ho01,Af05}
\begin{equation}\label{Lz0}
\frac{\langle L_z \rangle}{\hbar} = \frac{\langle r^2\rangle}{d_0^2} -1. 
\end{equation}
 For an anisotropic condensate, an analogous  result  follows from the expectation value $\langle L_z \rangle = \langle xp_y-yp_x \rangle$ in a LLL state.  Equations (\ref{xy}) for $x$ and $y$  and the corresponding relations for  $p_x$ and $p_y$ lead to an expression involving the combinations $\langle \alpha_+^\dagger\alpha_+ + \alpha_+\alpha_+^\dagger \rangle $ and $\langle( \alpha_+^\dagger)^2  + \left(\alpha_+\right)^2 \rangle$.  Comparison with Eqs.~(\ref{x2}) and (\ref{y2}) and use of Eq.~(\ref{gam}) then yields the following generalization of (\ref{Lz0})
\begin{equation}\label{Lz}
\frac{\langle L_z \rangle}{\hbar} = \frac{m\gamma}{2\hbar}\langle x^2 + y^2 \rangle +\frac{m\omega_-}{2\hbar}\left(\beta_- - \frac{1}{\beta_-} \right)\langle x^2 -y^2\rangle -\frac{1}{2}\left(\beta_- + \frac{1}{\beta_-}\right).
\end{equation}
For a symmetric trap ($\omega_x=\omega_y=\omega_0$), it follows by inspection that this result has the correct limit (\ref{Lz0}), because $\gamma\to 2\omega_0$,  $\langle y^2\rangle = \langle  x^2\rangle $, and $\beta_- \to1$.

\section{interacting gas in  a rotating anisotropic trap}

 A dilute interacting Bose-Einstein condensate with $N$ particles  in a trap is described by a condensate wave function $\psi$ that is here normalized to unity,  with $\int dxdy\, |\psi|^2 = 1$.   The Gross-Pitaevskii  (GP) energy functional for this system involves both the noninteracting Hamiltonian $H_0$ from Eq.~(\ref{h0}) and the interaction energy
 \begin{equation}\label{EGP}
E[\psi] = \int dxdy\left(\psi^* H_0\psi + \textstyle{\frac{1}{2}} g_{\rm 2d} N |\psi|^4\right),
\end{equation}
where $g_{\rm 2d}$ is a two-dimensional coupling constant with dimensions of energy times area.  If the condensate is confined in a tight axial harmonic trap with oscillator length $d_z = \sqrt{\hbar/(m\omega_z)}$, then
 $g_{\rm 2d} = \sqrt{8\pi}\,\hbar^2 a_s/(md_z)$, where $a_s$ is the $s$-wave scattering length~\cite{Fe03a,Af05,Wa06}.  In contrast, for a condensate that is  uniform in the $z$ direction with axial length $Z$, the analogous relation is $g_{\rm 2d} = 4\pi\hbar^2 a_s/(mZ)$.
The Euler-Lagrange equation for the wave function is the  stationary GP equation 
\begin{equation}\label{GP}
H_0\psi + g_{\rm 2d}N|\psi|^2 \psi = \mu \psi,
\end{equation}
where the chemical potential $\mu$ is fixed by the normalization of $\psi$.

\subsection{Lowest Landau-level  limit for rapid rotation}
 If the trap rotates rapidly, the condensate wave function $\psi$ can be approximated by a general LLL state $\psi_{LLL}=\sum_n c_n\varphi_{n0}$.   In this case, the expectation value of $H_0$ simplifies considerably to Eq.~(\ref{h0LLLa}).  Correspondingly, the total energy functional takes the approximate form 
 \begin{eqnarray}\label{ELLL}
E_{LLL}[\psi] & = & \frac{1}{2} \hbar\omega_- - \frac{1}{4}\hbar \omega_+ \left(\beta_+\beta_- + \frac{1}{\beta_+\beta_-}\right)  \nonumber \\[.2cm]
&  & +\int dxdy \left[ \frac{1}{2} m\gamma \omega_+\left(\beta_+x^2  + \frac{1}{\beta_+}  y^2  \right)|\psi|^2
+ \frac{1}{2} g_{\rm 2d} N |\psi|^4\right],
\end{eqnarray}
where I now follow~\cite{Af05} and  omit the subscript $LLL$  on the condensate wave function.

If the restriction to the LLL states is ignored, the absolute minimum of $E_{LLL}[\psi]$ is found by varying $|\psi|^2$ subject solely to the normalization condition.  The resulting  approximate GP equation 
\begin{equation}\label{LLLGP}
\frac{1}{2} m\gamma \omega_+\beta_+ x^2 + \frac{1}{2} m\gamma \omega_+\frac{1}{\beta_+}y^2 + g_{\rm 2d} N |\psi|^2 = \mu
\end{equation}
implies an anisotropic  density distribution 
\begin{equation}\label{TF}
|\psi_{\rm min}(x,y)|^2 = \frac{\mu}{g_{\rm 2d} N} \left(1-\frac{x^2}{R_x^2}  - \frac{y^2}{R_y^2}\right),
\end{equation}
with characteristic condensate radii given by 
\begin{equation}\label{RTF}
R_x^2 = \frac{2\mu }{m\gamma\omega_+\beta_+}, \qquad R_y^2 = \frac{2\mu \beta_+}{m\gamma\omega_+}.
\end{equation}
Note that  the ratio $R_y/R_x = \beta_+\propto \omega_+$ vanishes as $\Omega \to \omega_x$, but the behavior of the individual condensate radii requires a study of the chemical potential $\mu$.

As emphasized by various authors~\cite{Ho01,Ba04,Co04,Wa04,Af05,Wa06}, this density is very similar to 
the familiar Thomas-Fermi form for a nonrotating  condensate in a stationary two-dimensional trap.    In that case, the repulsive interactions expand the condensate and reduce the kinetic energy  compared to the trap energy and the interaction energy.  The  situation here is very different, because the approximate LLL wave function explicitly incorporates the full single-particle Hamiltonian, including the kinetic energy;  in this context, the appearance of the squared coordinates $x^2$ and $y^2$ arises from the special properties of the LLL states, specifically the result in Eq.~(\ref{h0LLLa}).

The normalization condition for $|\psi|^2$ in (\ref{TF}) readily yields the condition 
\begin{equation}\label{mu}
\mu = \sqrt \frac{m\gamma\, \omega_+\, g_{\rm 2d}N}{\pi}.
\end{equation} 
 Note that  the chemical potential $\mu$ vanishes proportional to $\sqrt{\omega_+} \propto \delta^{1/4}$  for a rapidly rotating anisotropic trap, where $\delta = 1-\Omega/\omega_x\to 0$.  Equation (\ref{TF}) then shows that the central density $|\psi_{\rm min}(0,0)|^2$ also has the same behavior in this limit.   

A combination of Eqs.~(\ref{RTF}) and 
(\ref{mu}) gives the condensate radii
\begin{equation}\label{RTFa}
R_x^2 = \frac{2}{\beta_+}\sqrt{\frac{g_{\rm 2d} N}{\pi m\gamma\,\omega_+}},\qquad R_y^2 = 2\beta_+ \sqrt{\frac{g_{\rm 2d} N}{\pi m\gamma\,\omega_+}}.
\end{equation}
 Correspondingly,  the normalized minimizing density (\ref{TF}) becomes 
\begin{equation}\label{TFmin}
|\psi_{\rm min}(x,y)|^2 = \frac{2}{\pi R_xR_y} \left(1-\frac{x^2}{R_x^2}  - \frac{y^2}{R_y^2}\right).
\end{equation}
Since $ \omega_+ $ and $\beta_+ $ are both proportional to $ \sqrt \delta$ as $\delta\to 0$ for fixed trap anisotropy, it is clear that $R_x^2$ grows like $\delta^{-3/4}$ and  $R_y^2$ shrinks like $\delta^{1/4}$ for small $\delta$, which reflects the conservation of total number of particles. In particular,  the total area $\pi R_xR_y$ diverges like $\omega_+^{-1/2}\propto \delta^{-1/4}$.  This anisotropy of the minimizing $N$-body condensate in the limit  $\delta\to0$ is quite different from the anisotropy of the  LLL single-particle ground state $\varphi_{00}$, where Eq.~(\ref{a}) shows that   $a_x$ grows but $a_y$ approaches a constant as $\delta\to 0$.
 For the minimizing density (\ref{TFmin}), it is straightforward to evaluate the mean-squared displacements $\langle x^2\rangle = \frac{1}{6}R_x^2$ and $\langle y^2\rangle = \frac{1}{6}R_y^2$.   The  mean angular momentum in Eq.~(\ref{Lz}) then becomes 
\begin{equation}
\frac{\langle L_z\rangle }{\hbar} = \frac{m}{12\hbar} \left(\omega_+\beta_+  + \omega_-\beta_-\right)  R_x^2 + \frac{m}{12 \hbar}\left(\frac{\omega_+}{\beta_+}  + \frac{\omega_-}{\beta_-} \right) R_y^2 - \frac{1}{2}\left(\beta_- + \frac{1}{\beta_-}\right).
\end{equation}

The condition $g_{\rm 2d} n(0)\ll \hbar\omega_-$ for the validity of the lowest Landau-level approximation can now be made explicit.  Use of  $g_{\rm 2d}n(0)\approx 2g_{\rm 2d}N/(\pi R_xR_y)$ from (\ref{TFmin}),  Eq.~(\ref{RTFa}) for the condensate radii and the relation $g_{\rm 2d} \approx 4\pi  \hbar^2 a_s/(mZ)$ for a uniform condensate of thickness $Z$  yields  
\begin{equation}\label{LLLcond}
2\sqrt{\frac{\gamma \omega_+}{\omega_-^2}\,\frac{Na_s}{Z}}\ll 1.
\end{equation}
 Since $\omega_+\to 0$ and $\omega_-$ remains nonzero for  sufficiently rapid rotation ($\Omega\to \omega_x$), this condition can always be satisfied.

  In the special case of a symmetric trap, the minimizing density has the isotropic form~\cite{Af05}  
 \begin{equation}
|\psi_{\rm min}(r)|^2 = \frac{2}{\pi R_0^2} \left( 1 - \frac{r^2}{R_0^2}\right) 
\end{equation}
with 
\begin{equation}\label{R0}
R_0^2 = \left[\frac{2g_{\rm 2d}N}{\pi m \omega_0(\omega_0-\Omega)}\right]^{1/2}.
\end{equation}
This  squared condensate  radius diverges as $\Omega\to \omega_0$.
Similarly, the absolute minimum of the LLL energy functional (\ref{ELLL}) for a symmetric trap becomes~\cite{Af05}  $\left. E_{LLL} \right|_{\rm min}  = \hbar\Omega+\frac{2}{3}m\omega_0^2 R_0^2$.  As a simple check, it is easy to verify that   $ -\left. \partial E_{LLL} \right|_{\rm min} /\partial\Omega = \hbar\left(\frac{1}{3}R_0^2/d_0^2-1\right)$;  this result agrees with   Eq.~(\ref{Lz0}) because $\langle r^2 \rangle = \langle x^2 +y^2\rangle = \frac{1}{3} R_0^2$ in the symmetric  limit.

\subsection{Density of vortices}

The general LLL state $\psi_{LLL}$ is a linear combination of the states $\varphi_{n0}$.  Apart from a normalization factor $1/\sqrt {n!}$, each of these states is  a polynomial $p_n(\zeta)$ multiplied by the ground state $\varphi_{00}$, where $\zeta$ from Eq.~(\ref{zeta}) is proportional to $x + i \beta_- y$.  Thus the general LLL state  also involves a polynomial in $\zeta$ that can formally be factorized to write 
\begin{equation}\label{LLL}
\psi_{LLL} \propto \varphi_{00}\,\prod_j \left(\zeta-\zeta_j\right) .
\end{equation}
The corresponding LLL particle density  $n_{LLL} = |\psi_{LLL}|^2$ becomes 
\begin{equation}
n_{LLL} \propto |\varphi_{00}|^2\,\prod_j|\zeta-\zeta_j |^2 .
\end{equation}
Apart from an additive constant, the logarithm of this relation gives~\cite{Ho01,Sh04,Sh04a,Af05}
\begin{eqnarray}
\sum_j\ln\left| (x-x_j)^2 + \beta_-^2(y-y_j)^2\right| &=& \frac{x^2}{a_x^2} +\frac{y^2}{a_y^2} + \ln n_{LLL}(\bm r)\nonumber\\[.2cm]
& = &  \frac{1}{a_x^2}\left(x^2 + \frac{\beta_- y^2}{\beta_+}\right) + \ln n_{LLL}(\bm r).
\end{eqnarray}
Here I use Eqs.~(\ref{a})  and (\ref{dens00}) for the anisotropic ground state and note that $a_y^2\, \beta_- = a_x^2\, \beta_+$.

To include the anisotropy of the complex variable $\zeta\propto x+i \beta_- y$, it is convenient to introduce the rescaled variables $x' = x$ and $y' = \beta_- y$.  Application of the rescaled Laplacian $\nabla'^2 = \partial^2 /\partial x'^2 + \partial^2 /\partial y'^2 $ readily gives
\begin{equation}
\sum_j \nabla'^2 \ln\left|\bm r'-\bm r_j'\right|^2 = \frac{2}{a_x^2} \left(1+ \frac{1}{\beta_+\beta_-}\right) +\nabla'^2 \ln n_{LLL}(\bm r).
\end{equation}
Since $\nabla'^2 \ln |\bm r'|^2 = 4\pi\, \delta(x')\delta(y') = (4\pi/\beta_-)\,  \delta(x) \delta(y) $, this relation implies
\begin{equation} \label{nv}
n_v(\bm r)  = \frac{m\gamma}{2\pi\hbar} +\frac{\beta_-}{4\pi} \,\nabla'^2 \ln n_{LLL} (\bm r), 
\end{equation}
where $n_v(\bm r)  = \sum_j\delta^{(2)} (\bm r-\bm r_j)$ is the two-dimensional vortex density, and I have again used Eq.~(\ref{a}).  For a rapid rotation speed $\Omega\lesssim \omega_x \le \omega_y$ and any trap anisotropy, the frequency  $\gamma = (\omega_-^2-\omega_+^2)/(2\Omega)$  is given in terms of elementary expressions from Eqs.~(\ref{op}) and (\ref{om}), as shown in Fig.~1(a) for a typical anisotropy ratio $\omega_y/\omega_x = 1.2$.

To estimate  the vortex density $n_v$ in Eq.~(\ref{nv}), assume that the equilibrium particle density $n_{LLL}$ is that given by  the  absolute minimum solution in Eq.~(\ref{TF}), with $n_{LLL}(\bm r)\propto 1-x^2/R_x^2-y^2/R_y^2$.  A straightforward calculation then yields 
\begin{eqnarray}\label{nvgen}
n_v(x,y) \approx \frac{m\gamma}{2\pi \hbar} %\nonumber \\[.2cm]
  &-&\frac{\beta_-}{2\pi\left(1-x^2/R_x^2-y^2/R_y^2\right)^2}\left[\frac{1}{R_x^2}  +\frac{1}{\beta_-^2R_y^2}  \right.\nonumber \\[.2cm]
 &&\quad\quad+ \left.\left(\frac{y^2}{R_y^2}-\frac{x^2}{R_x^2}\right)\left(\frac{1}{\beta_-^2R_y^2}-\frac{1}{R_x^2}\right)\right].
\end{eqnarray}
For an isotropic trap ($\omega_x = \omega_y = \omega_0$), this expression reduces to the well-known axisymmetric result~\cite{Wa04,Sh04,Sh04a}
\begin{equation}\label{nvsymm}
n_v(r) \approx \frac{m\omega_0}{\pi\hbar} - \frac{1}{\pi R_0^2}\,\frac{1}{\left(1-r^2/R_0^2\right)^2},
\end{equation}
where $R_0^2$ is given in Eq.~(\ref{R0}).

Comparison of these two expressions shows some interesting differences:  
\begin{enumerate}
\item For fixed angular velocity $\Omega\lesssim \omega_0$, the vortex density (\ref{nvsymm}) in a rapidly rotating symmetric trap decreases gradually and isotropically away from the center of the condensate.  Such behavior has been observed at lower angular velocities in the mean-field Thomas-Fermi regime~\cite{Cod04}.  In contrast, the general expression for the vortex density in Eq.~(\ref{nvgen}) displays explicit anisotropy between $x$ and $y$. 
\item To sharpen this analysis, it is convenient to focus on the central vortex density 
\begin{equation}\label{nv0}
n_v(0) \approx  \frac{m\gamma}{2\pi \hbar} 
  -\frac{1}{2\pi}\left(\frac{\beta_-}{R_x^2}  +\frac{1}{\beta_-R_y^2} \right).
\end{equation}
For a symmetric trap, Eq.~(\ref{nvsymm}) shows that $n_v(0)$ increases monotonically with increasing $\Omega\le \omega_0$, because the mean condensate radius $R_0$ grows in the same limit.  For an anisotropic trap, in contrast, the frequency $\gamma$ and $R_x^{-2}$ both decrease with increasing $\Omega$, whereas $R_y^{-2}$ increases.  Thus the combined effect of anisotropy and rapid rotation can, in principle, yield a central  vortex density $n_v(0)$ that varies  non-monotonically  with increasing $\Omega$, as can be seen  for typical numerical examples.
\end{enumerate}
 It is not clear whether either of these behaviors would be observable in practice.

The present discussion has focused on the  ``macroscopic'' parabolic density profile that provides an absolute minimum of the energy in the rotating frame, ignoring the local distortions associated with the  vortex cores.  In practice, these phenomena have very different length scales:  in the rapidly rotating limit, the vortex core and the intervortex separation are both of order $\sqrt{\hbar/(m\Omega)}$, whereas the condensate radii $R_x$ and $R_y$ are generally much larger.  Thus it is possible to  treat the condensate density as locally uniform over the size of an individual vortex.  If the vortex lattice is treated as  triangular and unbounded, this analysis yields a simple renormalization~\cite{Si02,Ba04,Wa04,Af05,Wa06} of the interaction constant $g_{\rm 2d} \to b g_{\rm 2d}$, where $b \approx 1.1596$ is the numerical value for a triangular Abrikosov vortex  lattice~\cite{Kl64}.  Apart from this rescaling of the interaction parameter, the  description remains essentially unchanged.

\section{conclusions and Discussion}

This work has examined the behavior of a  two-dimensional Bose-Einstein condensate in an anisotropic harmonic trap (with general trap frequencies $\omega_x\le \omega_y$) that rotates rapidly at an angular velocity $\Omega\lesssim \omega_x$.  The single-particle Hamiltonian in Eq.~(\ref{h0}) is exactly soluble~\cite{Va56,Li01,Ok04}, although the  detailed form  of the  low-lying  quantum-mechanical states  in Eq.~(30) for an arbitrary anisotropy has apparently not appeared previously.  

The ground-state wave function $\varphi_{00}(x,y)$ in Eq.~(\ref{phi00}) is an anisotropic Gaussian with $\Omega$-dependent characteristic lengths $a_x$ and $a_y$  given in (\ref{a}).  In addition, the ground state has a phase  proportional to $xy$~\cite{Fe74,Pi03a}. In the quantum problem, this behavior reflects   the classical velocity potential  for an irrotational (vortex-free) fluid confined in a rotating elliptical boundary~\cite{La45}.  

Similar to  the case of  a rapidly rotating symmetric condensate, the two eigenfrequencies $\omega_\pm$ in Eq.~(\ref{opm}) for a rapidly rotating  anisotropic condensate have very different magnitudes:  $\omega_+$ vanishes  as $\Omega \to \omega_x$ (the smaller of the two trap frequencies), but $\omega_-$ remains finite in this same limit.  If the mean interaction energy $\sim g_{\rm 2d} n(0)$ is small compared to the gap $\hbar\omega_-$ between the ground state and  the first excited Landau level, then the system can be described with the set of lowest Landau-level states $\varphi_{n0}(x,y)$, where $n$ is a non-negative integer describing the number of plus quanta, each with the small energy $\hbar\omega_+$. Apart from the Gaussian factor $\varphi_{00}(x,y)$,  these states  involve a polynomial $p_n(\zeta)$ of order $n$, where $\zeta$ single complex variable proportional to $x+ i\beta_- y$ and $\beta_-\le 1 $ is a positive constant for any  $0< \Omega\le \omega_x$.

Within the set of lowest Landau-level states $\psi_{LLL} = \sum_n c_n\varphi_{n0}$, the expectation value $\langle H_0\rangle$ of the single-particle Hamiltonian can be reduced to an anisotropic linear combination of $\langle x^2\rangle$ and $\langle y^2 \rangle$.  An anisotropic  Thomas-Fermi-like density  $|\psi_{\rm min}(x,y)|^2 \propto 1-x^2/R_x^2-y^2/R_y^2$ provides an absolute lower bound for the  total energy of the interacting system. Here the condensate radii $R_x$ and $R_y$ are given in Eq.~(\ref{RTFa});  $R_x^2$ grows and $R_y^2$ shrinks as $\Omega\to \omega_x$, while the area $\pi R_xR_y$ of the elliptical condensate  grows  slowly in the same limit.  

As emphasized by Ho~\cite{Ho01}, the particle density $n_{LLL}(x,y) = |\psi_{LLL}(x,y)|^2$ in the lowest Landau-level limit also contains a description of the associated vortex density $n_v(x,y)$. This situation arises because the general linear combination of lowest Landau-level states is essentially a polynomial $P(\zeta)$ in the single complex variable $\zeta \propto x + i \beta_- y$, and the zeros $\{\zeta_j\}$  of $P(\zeta)$ represent the positions of the vortices in the $xy $ plane.  The actual particle density $n_{LLL}$ has small-scale structure arising from the vortex cores, superposed on the parabolic global shape.  If this fine-grain aspect is ignored, the resulting vortex density follows immediately in Eq.~(\ref{nvgen}). 

This work raises several interesting questions:
\begin{enumerate}
\item   Feynman's familiar expression for the mean vortex density $n_F = m\Omega/(\pi \hbar)$ in a large symmetric rotating condensate  requires modification for an anisotropic rotating condensate because of the irrotational flow induced by the rotating walls.  Specifically, this irrotational flow contributes to the total angular momentum and thus lowers the energy $E' = E - \Omega L_z$  in the rotating frame,  delaying the first  transition to a state containing a single  quantized vortex~\cite{Fe74}.  Indeed,   the measured critical angular velocity $\Omega_c$ for the appearance of the first vortex in a rigid elliptical or rectangular cylinder containing  uniform superfluid $^{4}$He exceeds that for a circular cylinder by a factor that increases with increasing anisotropy~\cite{De75}, confirming the  theoretical prediction.  A similar but more complicated situation occurs for slowly rotating weakly interacting anisotropic Bose-Einstein condensates~\cite{Li01}.  Although the analogous situation with many vortices has not been studied in detail, the increased critical angular velocity for the appearance of the first vortex suggests that the mean vortex density  at moderate rotation speeds in an anisotropic trap is likely to be smaller than in a corresponding symmetric trap.

In contrast, for a rapidly rotating  anisotropic condensate, Eq.~(\ref{nv0}) gives the central density $n_v(0) \approx m\gamma/(2\pi \hbar)$, apart from finite-size corrections associated with  the condensate radii.  This value of $n_v(0)$  typically exceeds $n_F$, because $\gamma/(2\Omega)\to (3\omega_x^2+\omega_y^2)/(4\omega_x^2) > 1$ in the limit $\Omega\to \omega_x$. The physical basis for such  enhanced vortex density  is not immediately obvious, and additional clarification would be desirable.
\item  The present LLL approach ignores the fine-grain structure of the vortex lattice (apart from the renormalization of the interaction parameter $g_{\rm 2d}$ by  the ``Abrikosov parameter'' $b\sim 1.16$)~\cite{Si02,Ba04,Wa04,Af05,Wa06}.   In the special case of a nearly symmetric trap rotating close to the limit of instability, Oktel~\cite{Ok04} finds that the vortex lattice remains almost exactly  triangular, with the principal lattice planes aligned with the direction of weak trap confinement.  It is not clear whether this situation holds for arbitrary anisotropy and less extreme rotation speeds.  
\item As $\Omega$ approaches the weak confining frequency $\omega_x$, the condensate becomes essentially one dimensional  and the vortices must then rearrange themselves to form a one-dimensional structure~\cite{Si05,Sa05}.  Ultimately, this behavior may be pre-empted by some sort of transition to a correlated (nonsuperfluid) state, as has been predicted for a symmetric trap~\cite{Co01}.   
\end{enumerate}
It will be interesting to investigate these various questions in detail.

 \acknowledgments{I am grateful to G.\ Bertsch for early guidance on the literature about rotating deformed nuclei and rotating anisotropic oscillator potentials.}

  \end{document}